\documentclass[aps,preprint,showpacs,nofootinbib,12pt]{revtex4}
\usepackage{graphicx}
\usepackage{epsfig}
\usepackage[dvips]{color}

\begin{document}

\title{The magnetic dipole transitions in the $(c\bar{b})$ binding system}

\author{ Hong-Wei Ke$^{1}$,  Guo-Li Wang$^{2}$,   Xue-Qian Li$^3$ and Chao-Hsi
Chang$^{4,5}$}

\affiliation{
$^{1}$ School of Science, Tianjin University, Tianjin 300072, China \\
$^2$ Department of Physics, Harbin Institute of Technology, Harbin 150001, China\\
$^{3}$ School of Physics, Nankai University, Tianjin 300071, China\\
$^{4}$ CCAST (World Laboratory), P.O.Box 8730, Beijing 100190, P.R. China.\\
$^5$ Institute of Theoretical Physics, Chinese Academy of Sciences,
Beijing 100190, China}

\begin{abstract}
\noindent The magnetic dipole transitions between the vector mesons
$B_c^*$ and their relevant pseudoscalar mesons $B_c$ ($B_c$,
$B_c^*$, $B_c(2S)$, $B_c^*(2S)$, $B_c(3S)$ and $B_c^*(3S)$ etc, the
binding states of $(c\bar{b})$ system) of the $B_c$ family  are
interesting. To see the `hyperfine' splitting due to spin-spin
interaction is an important topic for understanding the spin-spin
interaction and the spectrum of the the $(c\bar{b})$ binding system.
The knowledge about the magnetic dipole transitions is also very
useful for identifying the vector boson $B_c^*$ mesons
experimentally, whose masses are just slightly above the masses of
their relevant pseudoscalar mesons $B_c$ accordingly. Considering
the possibility to observe the vector mesons via the transitions at
$Z^0$ factory and the potentially usages of the theoretical estimate
on the transitions, we fucus our efforts on calculating the magnetic
dipole transitions, i.e. precisely to calculate the rates for the
transitions such as decays $B_c^*\rightarrow B_c\gamma$ and
$B_c^*\rightarrow B_c\,e^+e^-$, and particularly work in the
Behte-Salpeter framework. In the estimate, as a typical example, we
carefully investigate the dependance of the rate
$\Gamma(B_c^*\rightarrow B_c\gamma)$ on the mass difference $\Delta
M=M_{B_c^*}-M_{B_c}$ as well.


\end{abstract}

\maketitle

\section{introduction}
Comparing with the hidden flavored heavy quarkonia such as charmonia
and bottomonia, the heavy meson $B_c$ and its family, being
explicitly double flavored, have not been thoroughly studied yet.
The reason is that not sufficient experimental data about $B_c$
meson are available and the experimental data about $B^*_c$ meson
(the lowest one) still are unavailable at all so far. $B_c$ and
$B_c^*$ are composed of two different heavy flavors, so that unlike
the production of the hidden flavored heavy quankonia, they cannot
be produced via a simple QCD process even at the hadron colliders.
At $e^+e^-$ colliders, the production is even more suppressed
because of absence of gluon fusion. The earlier work \cite{Chang}
indicates that one cannot expect to find $B_c$ in the cases with a
luminosity and collision energy as that of the LEP-I and II owing to
the production rate is small. As estimated by the authors of
ref.\cite{Chang}, the meson $B_c$ was first observed at a hadronic
collider, TEVATRON \cite{Cheung:1999ir}. It is natural that one
would expect to make a detailed study on the $B_c$ family at the
LHC, because the available energy and luminosity are much higher
than that of TEVATRON and the $B_c$-involved events should be
thousand times more. However, the messy QCD background of the hadron
colliders and the fact, that one cannot control the total longitude
momentum of the hadronic production, would contaminate the
environment and make precise measurements on $B_c$ very difficult,
and the observation on the other members of the $B_c$ family, i.e.
the excited states of $B_c$, almost impossible. In this aspect, the
proposed $Z^0$ factory possesses obvious advantage over the hadronic
colliders.

$Z^0$ factory, an $e^+-e^-$ collider with sufficiently high
luminosity and running at $Z^0$-boson pole, now is, as a phase of
ILC or independently, considered seriously. Even though the
inclusive production $e^+e^-\to B_c(\bar{B_c})+\cdots$ where two
pairs of heavy quarks ($c\bar c$ and $b\bar b$) emerges from a hard
gluon emission is suppressed, the high luminosity and the $Z$-pole
effects would greatly enhance the event-accumulation rate so that
the $B_c$ meson and its excited states (the other members of its
family) may be expected to be observed. Thus if the luminosity is
really high enough so the mesons $B_c$ and $B_c^*$ may be produced
numerously, thus the magnetic dipole radiative transitions may be
used to recognize the production of the excited states. In fact,
LEP-I did search for the $B_c$ meson and could not make any definite
conclusion, such as that the $B_c$ meson has been observed, due to
`low luminosity, so the small statistics for the events
\cite{LEP-I}.

$B_c$ being the ground state, its decay characteristics are
completely distinct from the hidden flavored heavy-quarkonia.
Namely, the $B_c$ can only decay via weak interaction, and its
lifetime has been carefully studied\cite{Chang1}. Whereas an excited
state of $B_c$ meson must decay to a lower excited or the ground
state via gluon (strong interaction) and/or photon (electromagnetic
interaction) emissions, and it depends on the quantum number and the
mass difference of the initial and final states. Moreover, it is
known that of the electromagnetic decays, the magnetic dipole $M1$
transitions between the vector and pseudoscalar states play an
important role.

In another work \cite{Ke:2009sk}, we discussed the possibility of
observing the radially excited states $B_c(ns)\; n>1$ via processes
$B_c(ns)\to B_c+\pi\pi$ at LHC and the $Z^0$ factory. Our
calculation is based on the QCD multi-pole expansion
method\cite{Gottfried} and we find that at the $Z^0$ factory, it
would be optimistic to observe the two-pion emission decays. On
other aspect, the nearest member to $B_c$ in the family is the
vector-boson $B_c^*\; (1s)$. With possible and precise spin-spin
interaction, one may estimate the splitting between $B_c(1s)$ and
$B^*_c(1s)$ explicitly as $30\sim50$ MeV, so that $B_c^*\to
B_c+\pi^0(\eta,\eta')$ is forbidden by the energy-momentum
conservation. Thus the dominant decay mode of $B_c^*$ would be the
magnetic dipole radiative decay $B_c^*\to B_c+\gamma$. The decay
$B_c^*\to B_c+e^+e^-$ is also governed by the electromagnetic
process and the products $e^+e^-$ would be easily caught by the
detector as a clear signal. Even though comparing with $B_c^*\to
B_c+\gamma$, its rate is suppressed by the three-body final phase
space and an extra electromagnetic vertex, its observation may still
be expected, because tracks of $e^+e^-$ would be easier to be
identified than that of a photon. Definitely, we can gain more
information about the $B_c^*$ and determine the mass splitting
$\Delta M=M_{B_c^*}-M_{B_c}$ from the data which will be available
at the $Z^0$ factory.

In Ref.\cite{Chang:2003ua} the authors explored radiative of
$M_1\rightarrow M_2\gamma$ in the Bathe-Salpeter(BS)
frameork\cite{BS}. Solving the BS equation one can obtain the wave
functions and eigenvalues of the bound state. With the Mandelstam
formula\cite{Mandelstam}, we calculate the transition matrix
elements between the bound states with appropriate BS wave
functions. Concretely, in terms of the formula given in
Ref.\cite{Chang:2003ua} we evaluate the transition matrix element of
$B_c^*\to B_c$ in the BS framework and extract the corresponding
form factor $F_{VP}(Q^2=0)$\cite{Choi:2007se}. With the form factor
we are able to calculate the rate of $B_c^*\to B_c+\gamma$. Then we
go on to evaluate the rate of $B_c^*\to B_c+e^+e^-$ where the photon
is virtual. In view of the progress in the experimental aspect we
also evaluate the transitions $B_c^*(2S)\to B_c+\gamma(e^+e^-)$,
$B_c^*(3S)\to B_c+\gamma(e^+e^-)$ and $B_c(2S)\to
B_c^*+\gamma(e^+e^-)$.

Our strategy is follows: first we solve the BS equation using the
parameters given in Ref.\cite{wangzhang} and get the spectra and
the wave functions of $B_c^*(nS)$ and $B_c(nS)$ respectively; then
with the formula obtained in Ref.\cite{Chang:2003ua}, we evaluate
the transition matrix element of $B_c^*(nS)\to B_c$, $B_c(2S)\to
B_c^*$, and $B_c(3S)\to B_c^*$ and $B_c(3S)\to B_c^*(2S)$ and
extract the form factors $F(Q^2=0)$; using these form factors the
rates of $B_c^*(nS)\to B_c+\gamma$, $B_c^*(nS)\to B_c+e^+e^-$,
$B_c(2S)\to B_c^*+\gamma, $ $B_c(2S)\to B_c^*+e^+e^-$, $B_c(3S)\to
B_c^*+\gamma, $ $B_c(3S)\to B_c^*+e^+e^-$, $B_c(3S)\to
B_c^*(2S)+\gamma, $ $B_c(3S)\to B_c^*(2S)+e^+e^-$ are eventually
obtained. After the introduction, we present the theoretical
formulae for calculating the rates of $V\to P+\gamma$ ($P\to
V+\gamma$) and $V\to P+e^+e^-$($P\to V+e^+e^-$), and then in
sec.III, we list our numerical results, the last section is
devoted to our discussion and conclusion.

\section{The formula of $V\rightarrow P$ and $P\rightarrow V$ in the  BS framework}
\subsection{$V\rightarrow P\,\gamma$}
In the BS framework the corresponding S-matrix element was
formulated as \cite{Chang:2003ua}
\begin{eqnarray}\label{2s1}
\langle
M_2(\mathcal{P}')\gamma(Q,\epsilon)|S|M_1(\mathcal{P})\rangle=\frac{(2\pi)^4e}{\sqrt{2^3\omega\omega_1\omega_2}}
\delta(\mathcal{P}'+Q-\mathcal{P})\epsilon_\mu \langle
M_2(\mathcal{P}')|J_{em}^\mu|M_1(\mathcal{P})\rangle,
\end{eqnarray}
where $M_1,\,M_2$ are the initial and daughter mesons,
$\mathcal{P}, \,\mathcal{P}'$ are their four-momenta and
$\omega_1, \omega_2$ are their energies. $Q$ is the momentum of
the emitted photon, $\epsilon$ is its polarization vector and
$\omega$ is its energy.

For the photon emission, the transition matrix element reads
\begin{eqnarray}\label{2s2}
 \langle
M_2(\mathcal{P}')|J_{em}^\mu|M_1(\mathcal{P})\rangle&=&\langle
M_2(\mathcal{P}')|J_{em}^\mu|M_1(\mathcal{P})\rangle_1+\langle
M_2(\mathcal{P}')|J_{em}^\mu|M_1(\mathcal{P})\rangle_2,\nonumber \\
&=& \int\frac{ d^3q_{\mathcal{P}\perp}}{(2\pi)^{3}} Tr\left\{Q_{1}
\frac{{\mathcal{P}}\!\!\!\!\slash}{m_{_{M_1}}}\left[
\bar{\varphi'}^{++}_{\mathcal{P}'}({q}_{\mathcal{P}\perp}+\alpha_2{\mathcal{P}'}_{\mathcal{P}\perp})
\gamma_{\mu}{\varphi}^{++}_{\mathcal{P}}({q}_{\mathcal{P}\perp})
\right]
\right.\nonumber \\
&&+ \left. Q_{2} \left[
\bar{\varphi'}^{++}_{\mathcal{P}'}({q}_{\mathcal{P}\perp}-\alpha_1{\mathcal{P}'}_{\mathcal{P}\perp})
\frac{{\mathcal{P}}\!\!\!\!\slash}{m_{_{M_1}}}{\varphi}^{++}_{\mathcal{P}}({q}_{\mathcal{P}\perp})
\right]\gamma_{\mu} \right\}\,.
\end{eqnarray}
where $Q_1(Q_2)$ is the charge carried by the quark(antiquark) and
other notations are listed in the Appendix. In the processes
$B_c^*(nS)\to B_c(n'S)+\gamma$ ($n\geq n'=1,2,\cdots$), $B_c^*(nS)$
is a vector($V$) and $B_c(n'S)$ is a pseudoscalar($P$). Due to the
quantum numbers of the initial and final states, the transitions
must be the nature of magnetic dipole, and there is only one form
factor for the current matrix elements, namely as that in
Ref.\cite{Choi:2007se} the general form factor $F_{VP}(Q^2)$ for
$V\rightarrow P\gamma^*$ is related to the current matrix element as
follows:
\begin{eqnarray}\label{2s3}
\langle
P(\mathcal{P}')|J_{em}^\mu|V(\mathcal{P})\rangle
=ie\epsilon^{\mu\nu\rho\sigma}\epsilon_\nu(\mathcal{P})Q_\rho
\mathcal{P}_\sigma F_{VP}(Q^2),
\end{eqnarray}
where  $Q=\mathcal{P}-\mathcal{P}'$ is the four momentum of the
virtual photon, $\epsilon_\nu(\mathcal{P})$ is the polarization
vector of the initial meson. $F_{VP}(Q^2)$ can be extracted by
evaluating the $\langle
M_2(\mathcal{P}')|J_\mu|M_1(\mathcal{P})\rangle$ in Eq.(\ref{2s2}).
For real photon case  $Q^2$ is equal to 0 i.e. $F_{VP}( Q^2=0)$.

\begin{figure}
\begin{center}
\begin{tabular}{cc}
\includegraphics[width=7cm]{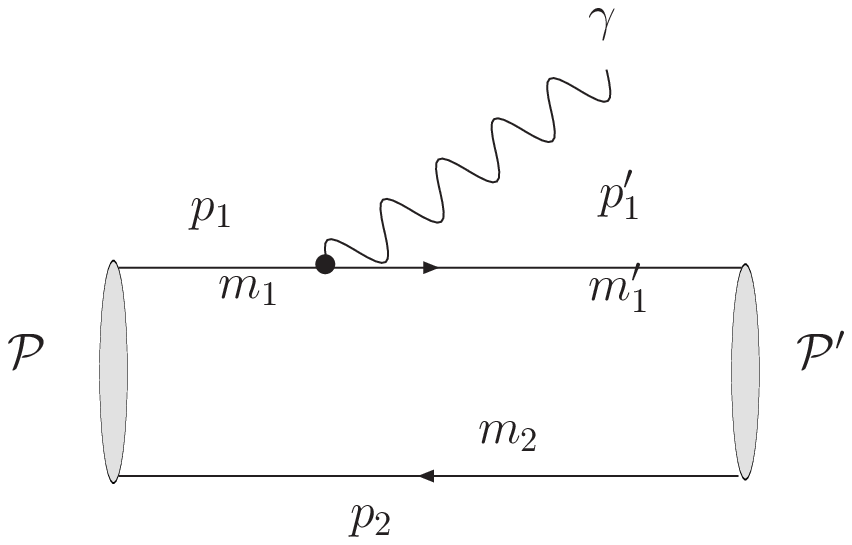}
\includegraphics[width=7cm]{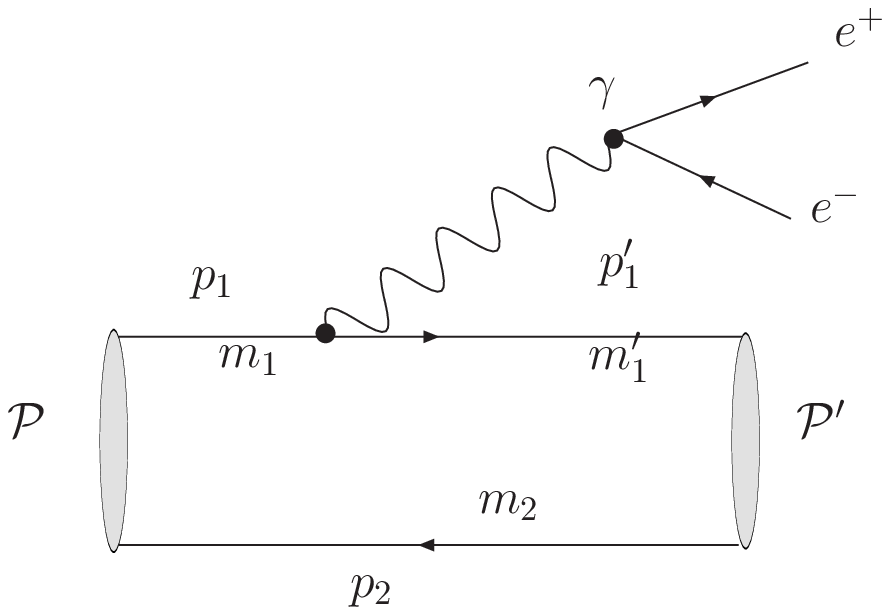}
\end{tabular}
\\+emission from antiquark
\\a\,\,\,\,\,\,\,\,\,\,\,\,\,\,\,\,\,\,\,\,\,\,\qquad\qquad\qquad\qquad\qquad\qquad\qquad\,b
\end{center}
\caption{The Feynman diagrams for the electromagnetic transition }
\label{fig:LFQM}
\end{figure}

The decay width of $V\rightarrow P\gamma$ is
\begin{eqnarray}\label{2s4}
\Gamma(V\rightarrow
P\gamma)=\frac{\alpha}{3}(\frac{m_V^2-m_P^2}{2m_V})^3F^2_{VP}(0),
\end{eqnarray}
where $\alpha$ is the fine-structure constant and $m_V,\;m_P$ are
the masses of the $B_c^*$ and $B_c$ respectively.

\subsection{$V\rightarrow P\,e^+e^- $}
The S-matrix element for $V\rightarrow P\,e^+e^- $ was given as
\begin{eqnarray}\label{2s5}
\langle
M_2(\mathcal{P}')e^+(p_a,s_a)e^-(p_b,s_b)|S|M_1(\mathcal{P})
\rangle&&=\nonumber\frac{(2\pi)^4e^2}{\sqrt{2^4\omega_a\omega_b\omega_1\omega_2}}
\delta(\mathcal{P}'+p_a+p_b-\mathcal{P})\frac{2m_e}{q^2}\\&&
\bar{U_e}(p_b,s_b)\gamma_\mu{U_e}(-p_a,s_a)\langle
M_2(\mathcal{P}')|J_{em}^\mu|M_1(\mathcal{P})\rangle,
\end{eqnarray}
where $p_a,\; p_b$ are the four-momenta of $e^+$ and $e^-$,
$\omega_a,\;\omega_b$ are their energies and $U_e,\;\bar{U}_e$ are
the corresponding spinors with spins $s_a$ and $s_b$. Using
Eq.(\ref{2s3}), $\langle
M_2(\mathcal{P}')|J_{em}^\mu|M_1(\mathcal{P})\rangle$ can be
parameterized into $F_{VP}(Q^2)$ which can be calculated according
to  Eq.(\ref{2s2}).

In terms of the formula given in Ref.\cite{PDG08} we obtain
\begin{eqnarray}\label{2s6}
d\Gamma=\frac{16\alpha^2\,F^2_{VP}(Q^2)\left( 11m_e^2 + 4Q^2
\right) {\sqrt{Q^4-4m_e^2\,Q^2  }}
    {\left[  {\left( -m_V^2 + m_P^2 + Q^2 \right) }^2  -4m_P^2Q^2 \right]
    }^{\frac{3}{2}}}{4608m_V^3\,
    \pi \,Q^6}dQ^2,
\end{eqnarray}
where $Q=p_1+p_2$. Integrating out $Q^2$ in the expression
(\ref{2s6}) one  obtains the decay width of $B_c^*\rightarrow
B_c\,e^+e^- $.

For the transition of $P\rightarrow V$, the form factor
$F_{PV}(Q^2)$ can be obtained as we did for $F_{VP}(Q^2)$. We still
can use Eq.(\ref{2s4})(Eq.(\ref{2s6})) to calculate the rate
$B_c(2S)\rightarrow B_c^* \gamma(e^+e^-)$ by replacing $F_{PV}(Q^2)$
with $F_{VP}(Q^2)$.

In principle, we can extend our computations to higher excited
states  and the P-wave states of the $B_c$ family, but because their
production rates are much lower and experimental measurements would
be much more difficult, we do not intend to include them in this
work.

\section{Numerical results}
By solving the corresponding BS equations for $B_c^*(nS)$  and
$B_c(nS)$, their  wave functions and masses were evaluated  in
Ref.\cite{wangzhang} where the authors systematically explored the
spectra of mesons made of only heavy flavor quark-antiquark  and
all the free parameters in the theoretical model were fixed by
fitting the data of heavy quarkonia and $B_c$. The masses of
$B_c^*$ , $B_c^*(2S)$, $B_c^*(3S)$ and $B_c(2s)$ are obtained as
6.3308 GeV, 6.9103 GeV, 7.2755 GeV and 6.8623 GeV respectively.
The forms of the BS wave functions for the vector and pseudoscalar
mesons are listed in the Appendix and the corresponding parameters
can be found in Ref.\cite{wangzhang}.

\begin{table}
\caption{ The theoretical predictions of the rates for several
electromagnetic decay modes} \label{tab:modes}
\begin{tabular}{c|ccccc}\hline\hline
 & $|F(Q^2=0)|$(GeV$^{-1}$)  &\qquad $\Gamma_{(M_1\rightarrow M_2 \,\gamma)}$(keV)&
$\qquad\Gamma_{(M_1\rightarrow M_2 \,e^+e^-)}$ (keV)
\\\hline
$B_c^*\rightarrow B_c$&0.208&17.1$\times 10^{-3}$&8.64$\times
10^{-5}$
\\ $B_c^*(2S)\rightarrow B_c$ &0.023&0.28&1.59$\times
10^{-3}$
\\ $B_c^*(3S)\rightarrow B_c$ &0.014&0.37&2.11$\times
10^{-3}$
\\ $B_c(2S)\rightarrow B_c^*$ &0.030&0.38&1.65$\times
10^{-3}$\\
$B_c(3S)\rightarrow B_c^*$ &0.0072&0.074&0.42$\times 10^{-3}$\\
$B_c(3S)\rightarrow B_c^*(2S)$ &0.049&0.25&1.41$\times
10^{-3}$\\\hline\hline
\end{tabular}
\end{table}

Using the  wave functions of the initial and daughter mesons, we
calculate the transition matrix element in Eq.(\ref{2s2}) and
extract the form factor as $|F_{VP}(Q^2=0)|=0.208$ GeV$^{-1}$.
Substituting the value of $|F_{VP}(Q^2=0)|$ into Eq.(\ref{2s4}) we
get the width $\Gamma(B_c^*\rightarrow B_c \gamma)=17.1\times
10^{-3}$ keV. In Ref.\cite{Choi:2009ai} the authors used light-front
quark model to study $B_c^*\rightarrow B_c \gamma$ and obtained
$\Gamma(B_c^*\rightarrow B_c \gamma)=22.4(19.9)\times 10^{-3}$ keV
for $\Delta M=50$ MeV which is consistent with our result.

\begin{figure}
\begin{center}
\begin{tabular}{cc}
\includegraphics[width=7cm]{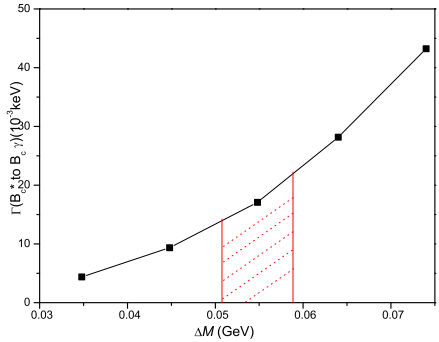}
\end{tabular}
\end{center}
\caption{ The dependance of $\Gamma(B_c^*\rightarrow B_c\gamma)$ to
$\Delta M$. The shadowed region is centered at the mass of
$M_{B_c^*}=6.3308$ GeV and the area corresponds to the experimental
errors which are taken as inputs to our numerical computation }
\label{fig:dm}
\end{figure}

$\Gamma(B_c^*\rightarrow B_c\gamma)$ is sensitive to the mass of
$m_{B_c^*}$ (or $\Delta M=m_{B_c^*}-m_{B_c}$ as $m_{B_c}$ has
already been experimentally determined) since the rate is
proportional to $\Delta M^3$.  Fig.\ref{fig:dm} shows the
dependence of $\Gamma(B_c^*\rightarrow B_c\gamma)$ on $\Delta M$.
In our calculation the form factor $|F_{VP}(Q^2=0)|$ hardly
changes for the different values of $m_{B_c^*}$ and it is nearly
equal to 0.208 GeV$^{-1}$.

With the wave functions of $B_c^*$ and $B_c$, $m_{B_c}=6.276$ GeV
we also can obtain the form factor $F_{VP}(Q^2)$ for $Q^2\neq 0$
$i.e.$ the emitted photon is an off-shell virtual one. For the
transition $B_c^*\rightarrow B_c\,e^+e^-$,  $Q^2$ varies from
$Q^2_{min}=(2m_e)^2$ to $Q^2_{max}=(m_{B_c^*}-m_{B_c})^2$.
We find the $|F_{VP}(Q^2)|$ is almost a constant for our
calculation accuracy, thus we set
$|F_{VP}(Q^2)|=|F_{VP}(Q^2_{min})|=0.208$ GeV$^{-1}$. Integrating
$d\Gamma$, we eventually obtain the width $\Gamma(B_c^*\rightarrow
B_c\,e^+e^- )=8.64\times 10^{-5}$ keV. The decay rates including
the transitions:  $B_c^*(2S)\rightarrow B_c$,
$B_c^*(3S)\rightarrow B_c$, $B_c(3S)\rightarrow B_c^*$ and
$B_c(3S)\rightarrow B_c^*(2S)$  are listed in Tab.\ref{tab:modes}.

\section{Conclusion}
The family of $B_c$ meson is composed by two different heavy
flavors: $b\bar c$, the members' production and decays are different
from those for hidden-flavored heavy quarkonia, and the study of the
$b\bar c$ system must be helpful in gaining insights into the hadron
structure and the governing physical mechanisms. The two heavy
quarks with different flavors cannot annihilate into gluons, in
addition to that the physics is rich, the influence of the
relativistic effects is alleviated, and in the BS framework the
instantaneous approximation seems to work well and the results are
more reliable. Even though the ground state, $B_c$ of $J^P=0^-$, was
found several years ago, its partner  $B_c^*$ of $J^P=1^-$ has not
been seen yet. Fortunately LHC begins running and a $Z^0$ factory is
proposed, both of them will offer us optimistic opportunity to
explore $b\bar c$ family, especially, $B_c^*$.

In this work we mainly study the transitions $B_c^*\rightarrow
B_c\gamma$ and $B_c^*\rightarrow B_c\, e^+ e^-$ in the BS
framework. Writing the transition matrix element in the form of
Eq.(\ref{2s2}), we determine the form factor $F_{VP}(Q^2)$. In the
calculation, we substitute the BS wave functions of initial and
daughter mesons which are obtained by solving the BS equation.
With the form factor $F_{VP}(Q^2)$, we evaluate
$\Gamma(B_c^*\rightarrow B_c\gamma)$ and $\Gamma(B_c^*\rightarrow
B_c\, e^+ e^-)$. When the mass of $m_{B_c^*}$ is 6.3308 GeV  and
$m_{B_c^*}=6.276\pm 0.004$ GeV (the measured value \cite{PDG08}
i.e. $\Delta M= 55\pm 4$ MeV), we obtain $\Gamma(B_c^*\rightarrow
B_c \gamma)=17.1\times 10^{-3}$ keV  and $\Gamma(B_c^*\rightarrow
B_c\,e^+e^- )=8.64\times 10^{-5}$ keV. The branching ratio of
$B_c^*\rightarrow B_c \gamma$ is three orders larger than that of
$B_c^*\rightarrow B_c\,e^+e^-$, it means that the chance of
observing $B_c^*\rightarrow B_c \gamma$ seems to be superior to
$B_c^*\rightarrow B_c\,e^+e^-$, however the positron and electron
are charged and their tracks would be easier caught by the
detector than a single photon, so that $B_c^*\rightarrow
B_c\,e^+e^-$ may still have its advantage for detection. We will
rely on the Monte-Carlo simulation made by our experimental
colleagues to make a judgement if at the $Z^0$ factory it is a
possible process to be measured.

Since the value of $B_c^*\rightarrow B_c \gamma$ is sensitive to the
vary of $\Delta M$ we study the dependance of
$\Gamma(B_c^*\rightarrow B_c \gamma)$to $\Delta M$. Our calculation
indicates that the value of $F_{VP}(Q^2)$ is not sensitive to the
$m_{B_c^*}$ but $\Gamma(B_c^*\rightarrow B_c\gamma)$ is rather
sensitive to the mass splitting $\Delta M$. We calculate the decay
width based on the BS framework, the obtained
$\Gamma(B_c^*\rightarrow B_c\gamma)$  is accordant with that
obtained in terms of the light-front-quark model by the authors of
Ref.\cite{Choi:2009ai}. In the work of Ref.\cite{Choi:2009ai}, the
authors used the variational method to fix the free parameters,
whereas we determine the parameters by fitting the data for
quarkonia. Application of both the light-front quark model and the
BS framework seems to be reasonable to deal with the radiative
process, however, the difference of the two theoretical evaluated
values may hint the feasibility of their application in this case.
Fortunately, the future experiments at the $Z^0$ factory will make a
more accurate measurement on $\Gamma(B_c^*\rightarrow B_c\gamma)$
and $\Gamma(B_c^*\rightarrow B_c+e^+e^-)$, and the data would judge
which model to be more reasonable. We are expecting the new data.

It is worth emphasizing again, even though at LHC, a large database
on $B_c$ and $B_c^*$ will be available, but the complicated
background makes a precise observation of $\Gamma(B_c^*\rightarrow
B_c\gamma)$ rather difficult, so we lay our hope on the proposed
$Z^0$ factory.

\section*{Acknowledgments}

This work is supported by  the special grant for new faculty from
Tianjin University. This work is partially supported by the National
Natural Science Foundation under the contract No. 10775073, No.
10875032, No.10875155, No.10847001 and the Special grant for the
PH.D program of the Education Ministry of China.

\appendix
\section{Notations}
Concerning  how to solve the BS equation the readers are suggested
to refer  Ref. \cite{Chang:2003ua,BSwang,wangzhang}. Here we only
present some notations appearing in this paper for readers'
convenience.

For a bound state of two constituents with the total momentum
$\mathcal{P}$ and relative momentum $q$,  $\mathcal{P}$ and $q$ are
defined as:
\begin{eqnarray*}
p_1=\alpha_1\mathcal{P}+q,\ \alpha_1=\frac{m_1}{m_1+m_2},\\
p_2=\alpha_2\mathcal{P}-q,\ \alpha_2=\frac{m_2}{m_1+m_2}.
\end{eqnarray*}

The relative momentum $q$ is divided into two parts,
$q_{_{\mathcal{P}_\|}}$ and $q_{_{\mathcal{P}_\bot}}$ and they are
longitudinal and transverse  to $\mathcal{P}$, respectively:
\begin{eqnarray}
q^{\mu}=q_{_{\mathcal{P}_\|}}^{\mu}+q_{_{\mathcal{P}_\bot}}^{\mu},
\end{eqnarray}
where $q_{_{\mathcal{P}_\|}}^{\mu}\equiv(\mathcal{P}\cdot
q/M^2)\mathcal{P}^{\mu}$,\ $q_{_{\mathcal{P}_\bot}}^{\mu}\equiv
q^{\mu}-q_{_{\mathcal{P}_\|}}^{\mu}$, and $M$ is the mass of the
bound state.

For the finial state with the total momentum $\mathcal{P}'$, the
momentum $\mathcal{P}'$ is also divided into two parts,
$\mathcal{P}'_{_{\mathcal{P}_\|}}$ and
$\mathcal{P}'_{_{\mathcal{P}_\bot}}$, longitudinal and transverse to
the momentum $\mathcal{P}$ of initial state, respectively:
\begin{eqnarray}
\mathcal{P}'^{\mu}={\mathcal{P}'}_{_{\mathcal{P}_\|}}^{\mu}+{\mathcal{P}'}_{_{\mathcal{P}_\bot}}^{\mu},
\end{eqnarray}
where
${\mathcal{P}'}_{_{\mathcal{P}_\|}}^{\mu}\equiv(\mathcal{P}\cdot
{\mathcal{P}'}/M^2)\mathcal{P}^{\mu}$,\
${\mathcal{P}'}_{_{\mathcal{P}_\bot}}^{\mu}\equiv
{\mathcal{P}'}^{\mu}-{\mathcal{P}'}_{_{\mathcal{P}_\|}}^{\mu}$.

Let us introduce several important notations:
\begin{eqnarray}\label{ab02}
&&\varphi_{_\mathcal{P}}^{\pm\pm}(q_{_{\mathcal{P}_\bot}}^{\mu})\equiv\Lambda_{1{_\mathcal{P}}}^{\pm}
(q_{_{\mathcal{P}_\bot}}^{\mu})\frac{{\mathcal{P}}\!\!\!\!\slash}{M}\varphi_{_\mathcal{P}}(q_{_{\mathcal{P}_\bot}}^{\mu})
\frac{{\mathcal{P}}\!\!\!\!\slash}{M}\Lambda_{2{_\mathcal{P}}}^{\pm}(q_{_{\mathcal{P}_\bot}}^{\mu}),\nonumber\\
&&{\bar\varphi}_{_\mathcal{P}}^{\pm\pm}(q_{_{\mathcal{P}_\bot}}^{\mu})\equiv
-\gamma_0\left[\varphi_{_\mathcal{P}}^{\pm\pm}(q_{_{\mathcal{P}_\bot}}^{\mu})\right]^\dagger
\gamma_0,
\end{eqnarray}
and
\begin{eqnarray}
\Lambda_{i{_\mathcal{P}}}^{\pm}(q_{_{\mathcal{P}_\bot}}^{\mu})&=&\frac{1}{2\omega_{i{_\mathcal{P}}}}
\left[\frac{{\mathcal{P}}\!\!\!\!\slash}{M}\omega_{i{_\mathcal{P}}}\pm
J(i)(m_i+{\not\!q}_{_{\mathcal{P}_\bot}})\right],\nonumber\\
\omega_{i{_\mathcal{P}}}&=&\sqrt{m_i^2+q_{_{\mathcal{P}_T}}^2},\,\,\,q_{_{\mathcal{P}_T}}=\sqrt{-q_{_{\mathcal{P}_\bot}}^2},
\end{eqnarray}
where $i$=1, 2 correspond to the quark and anti-quark, respectively,
and $J(i)=(-1)^{i+1}$.

The relativistic wave function for the  mesons with the quantum
numbers $J^P=0^-$ and $J^P=1^-$ can be generally written as
\begin{eqnarray}\label{aa01}
\varphi_{0^-}(q_{_{\mathcal{P}_\bot}})&&=\Big[f_1(q_{_{\mathcal{P}_\bot}}){\mathcal{P}}\!\!\!\!\slash+f_2(q_{_{\mathcal{P}_\bot}})M+
f_3(q_{_{\mathcal{P}_\bot}})\not\!{q_{_{\mathcal{P}_\bot}}}+f_4(q_{_{\mathcal{P}_\bot}})\frac{{\mathcal{P}\!\!\!\!\slash}\not\!{q_{_{\mathcal{P}_\bot}}}}{M}\Big]\gamma_5,\nonumber\\
\varphi_{1^-}^\lambda(q_{_{\mathcal{P}_\bot}})&=&q_{_{\mathcal{P}_\bot}}\cdot
\epsilon_{_\bot}^\lambda\left[g_1(q_{_{\mathcal{P}_\bot}})+g_2(q_{_{\mathcal{P}_\bot}})
\frac{\mathcal{P}\!\!\!\!\slash}{M}\right.\left.+g_3(q_{_{\mathcal{P}_\bot}})\frac{\not\!{q_{_{\mathcal{P}_\bot}}}}{M}
+g_4(q_{_{\mathcal{P}_\bot}})\frac{\mathcal{P}\!\!\!\!\slash\not\!{q_{\mathcal{P}_\bot}}}
{M^2}\right]\nonumber\\
&&+gf_5(q_{_{\mathcal{P}_\bot}})M\not\!\epsilon_{_\bot}^\lambda
+g_6(q_{_{\mathcal{P}_\bot}})\not\!\epsilon_{_\bot}^\lambda\mathcal{P}\!\!\!\!\slash+g_7(q_{_{\mathcal{P}_\bot}})(\not\!{q_{_{\mathcal{P}_\bot}}}{\not\!\epsilon}_{_\bot}^\lambda
-q_{_{\mathcal{P}_\bot}}\cdot\epsilon_{_\bot}^\lambda)\nonumber\\
&&+g_8(q_{_{\mathcal{P}_\bot}})\frac{(\mathcal{P}\!\!\!\!\slash\not\!\epsilon_{_\bot}^\lambda\not\!{q_{_{\mathcal{P}_\bot}}}
-\mathcal{P}\!\!\!\!\slash
q_{_{P_\bot}}\cdot\epsilon_{_\bot}^\lambda)}{M},
\end{eqnarray}
where $f_i$($g_i$) are scalar functions and can be obtained by
solving the BS equation.

\end{document}